# An Application of Ubiquitous Video Services and Management Systems in Civil Defense

## Zhaoming Dai

Xiaofeng Environmental Protection Technology, China

**Abstract** This paper proposes a model of ubiquitous video services and management systems for solving current issues of video surveillance services and management in the smart city. The author presents the Enterprise Service Bus (ESB) based methods of integrating and using existing video resources in civil defense emergency commanding by providing a use case "Civil Defense Emergency Command Center in Huai-An City", focusing on ubiquitous video services and management. Furthermore, this paper presents the differentiated services (DiffServ) scheduling algorithm for realizing the coordination of multitasking load from the prospective of providing various services aims at accelerating emergency responding, providing solutions to meet the needs of various clients for the video services, and making useful attempt to make contribution to the information construction of smart city in the future.

**Keywords**　　Video surveillance, Service oriented architecture, Enterprise service bus, Differentiated services scheduling

## 泛在视频服务与管理系统在民防的应用

**摘　要**　　本文针对智慧城市视频监控统一服务和管理方面存在的问题，提出了泛在视频服务与管理系统模型，以 ESB 为基础，以泛在公共视频服务与管理为目标，结合"淮安市民防应急指挥系统"应用实例，介绍了在民防应急指挥中充分利用和整合现有视频资源的方法，并从提供区分服务角度出发，提出了实现多事务负载均衡调度算法，为提高事务响应速度，满足各类用户对视频服务并发请求的需求提供了有效的解决方案，也为今后智慧城市信息化建设做了些有益的尝试。

**关键词**　　视频监控；SOA 企业服务总线 ESB 区分服务调度

## 0　引　言

随着城市现代化建设的飞速发展，城市社会监管面临着应对治安和重大突发公共事件的巨大压力，促使城市视频监控系统建的设规模愈来愈大。随之而来的问题是由于缺少顶层设计和组织协调，各个视频监控系统的建设自成体系，不能通过网络实现有效的视频资源共享，也不能形成一个完整的监控综合应用体系，制约了技术防范在城市社会监管中所发挥的作用，同时也难以满足视频监控专业化、社会化服务的要求。

泛在（Ubiquitous）视频服务与管理系统就是希望通过企业服务总线（Enterprise Service Bus，ESB）技术将各个自成体系的视频监控系统整合起来，实现应用层的数据、业务流程、信息和应用的高度融合，使视频服务无时不在、无处不在[1-4]。

本文提出了一个基于 ESB 思想的泛在视频服务与管理系统，系统通过 ESB 消息共享机制，实现了网络应用程序间数据共享，使异构环境下的各个系统采集的视频得以泛在应用，并以淮安市民防应急指挥系统应用为例，具体阐述系统的设计、实现和关键技术。

## 1　系统模型

要使集成异构环境下的视频监控系统能在Internet上提供无时不在、无处不在的实时公共视频服务，主要困难是如何将在不同操作系统、不同通信协议、不同视频格式，以及不同访问方式开发的视频系统整合起来，使其数据能够共享，应用能够融合。这也是信息系统集成领域自手工业务集成向企业应用集成（Enterprise Application Integration，EAI）进化，再到面向服务架构（Service-Oriented Architecture，SOA），直至当今的以ESB为基础的SOA技术[5-8]。

ESB是构建基于面向SOA解决方案时所使用基础架构的关键部分，是由中间件技术实现并支持SOA的一组基础架构功能[9-14]。ESB在实施SOA的过程中扮演重要的基础构件角色，通过ESB，将原有应用系统功能以服务的形式连接在一起，从而实现已有系统与新系统的无缝整合[15-19]。

泛在视频服务系统采用SOA技术，能够灵活的在

数据、流程、应用等各层面上实现应用系统整合，将相互独立的异构环境下视频监控系统资源进行统一服务与管理，实现视频资源的注册、发现、安全和访问等服务功能，并具有用户优先级、视频资源更新等管理功能，从而提供一种泛在的公共视频服务[20,21]。系统以服务网站、客户端和标准服务接口等多种方式，来满足不用类型使用者的应用需求。

系统模型如图1所示

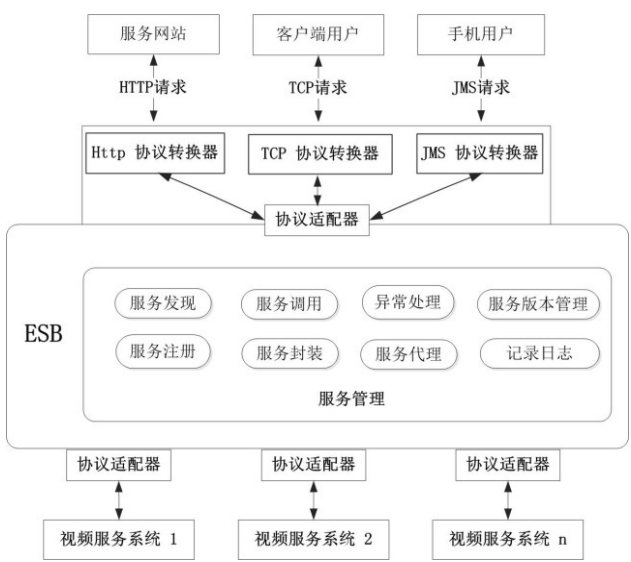

图1 系统模型图

## 2 应用实例

淮安市民防应急指挥系统主要功能是当突发公共事件发生时，系统能提供应急指挥调度时所需的数据、视频、语音等信息，并实现指挥中心与机动指挥所、指挥中心与单兵、指挥中心与现场救援队伍的实时音视频交互。应急指挥系统所需的音视频信息一般是通过已安装在城市中的视频监控系统和移动视频系统获取，如治安视频监控系统、道路交通视频监控系统、城管视频监控系统、安监视频监控系统、机动指挥所视频服务系统，以及由服务提供商提供的音视频监控系统，如"全球眼"、"千里眼"等。民防应急指挥系统就是要能快捷、灵活和充分利用全市的各类视频资源。

泛在视频服务与管理系统针对不同厂家音视频系统调用方法、访问参数和视频编解码都各不相同的难题，采用 Web Service 概念模型，屏蔽底层实现和存储机制，将应用系统封装，直接调用所描述的接口。在底层，利用 ESB 技术，通过适配器实现与其它组件互联互通，满足异构系统集成的需求。

开源的 Mule ESB 是目前世界上使用最为广泛的企业服务总线和集成平台之一，它是为支持不同系统和服务之间的高性能、多协议业务而设计的，泛在视频服务与管理系统选用 Mule 作为 ESB 的实现平台，提供了一个基于事务处理的分布式架构，可实现多个异构系统服务之间数据交互[22]。

本系统将多个 Mule 集成到应用环境中，构成了一个 Mule 服务骨干（Mule Services Backbone）网络架构，实现异构视频监控系统的数据和服务的获取，为应用提供各种类型的交互与共享。

### 2.1 系统结构

人防应急指挥系统结构如图 2 所示。

系统总体结构分为三层：底层是各种异构环境下各厂商视频服务系统，中间层是基于 Mule ESB 的泛在视频服务平台，最上层是为各类用户提供视频服务与管理的应用层。

泛在视频服务平台作为中间层，通过各类协议适配器，将不同的视频监控系统接入平台。访问视频服务系统的调用参数加密存储到系统平台的数据库中，管理系统能够动态检测视频服务系统参数变化，自动更新视频服务系统参数[23]。平台同时为用户提供多种协议服务端口，如 HTTP、TCP、JMS 等，实现了一致、透明的泛在视频服务与管理的应用。

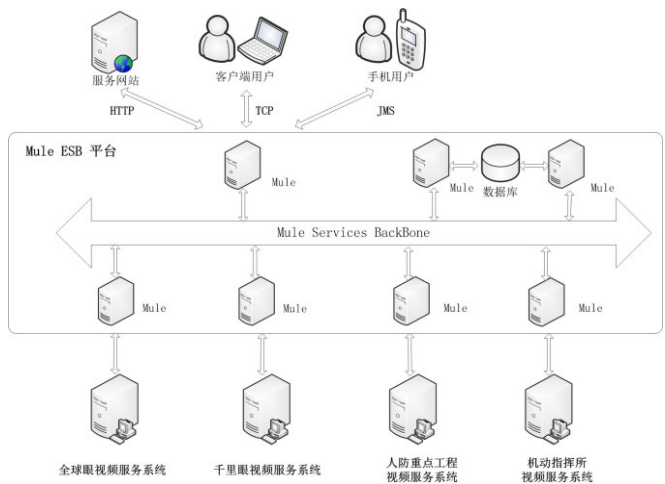

图 2 系统结构图

### 2.2 服务流程

泛在视频服务系统主要包括视频资源接入、视频资源检索、视频系统状态监测、安全管理、用户管理、视频资源管理和服务管理等功能，以下着重介绍视频资源检索的服务流程。

视频资源检索是本系统的核心服务，针对用户可在不同环境下使用本系统服务的特点，本系统设计了三种类型的服务访问协议接口，以同时满足 HTTP、TCP、JMS 用户的调用。下面以用户通过系统提供的网站页面使用音视频服务为例，介绍音视频资源检索服务的处理流程。

## 2.2.1 用户身份验证

**Keywords** 用户通过系统网站的登录界面登录系统。

**Keywords** 系统接收用户提交身份信息,调用身份验证服务,对用户身份进行认证,配置文件主要代码如下:

//系统使用 HTTP 端点监听并接收用户通过服务网站提交的身份信息。

```
<http:inbound-endpoint
exchange-pattern="request-response"
doc:name="HTTP"
address="http://www.videoserver.com/user_login"/>
```

//创建 VM 端点,将用户身份信息提交验证

```
<vm:outbound-endpoint
exchange-pattern="request-response" doc:name="VM"/>
```

**Keywords** 身份验证服务通过 VM 端点,获得用户身份信息,使用 JDBC 实现数据库连接,并动态执行 sql 的语句,进行用户身份认证。配置文件关键代码如下描述:

//创建 VM EndPoints

```
<vm:connector name="vmQueue"/>
<endpoint address="vm://query" name="query" exchange-pattern="one-way"/>
```

//建立 spring 数据源

```
<spring:bean id="dataSource"
class="org.apache.commons.dbcp.BasicDataSource"
destroy-method="close">
    <spring:property name="driverClassName"
value="com.mysql.jdbc.Driver" />
<spring:propertyname="url"
value="jdbc:mysql://localhost/VideoServerDB" />
  <spring:property name="username" value="admin" />
  <spring:property name="password" value="mysqlP" />
  <spring:property name="maxActive" value="50" />
  <spring:property name="maxIdle" value="10" />
  <spring:property name="maxWait" value="1000" />
  <spring:property name="defaultAutoCommit"
value="true" />
  </spring:bean>
```

// 创建 SQL select 语句

```
<jdbc:connector name="jdbcConnector"
dataSource-ref="dataSource">
<jdbc:query key="selectUser" value="SELECT
username,usertype FROM app_user where
username=#[map-payload:username]" />
</jdbc:connector>
```

**Keywords** 然后将验证结果再传输给 HTTP 端点,在登录界面上显示出验证结果,主要流程如图 3 所示。

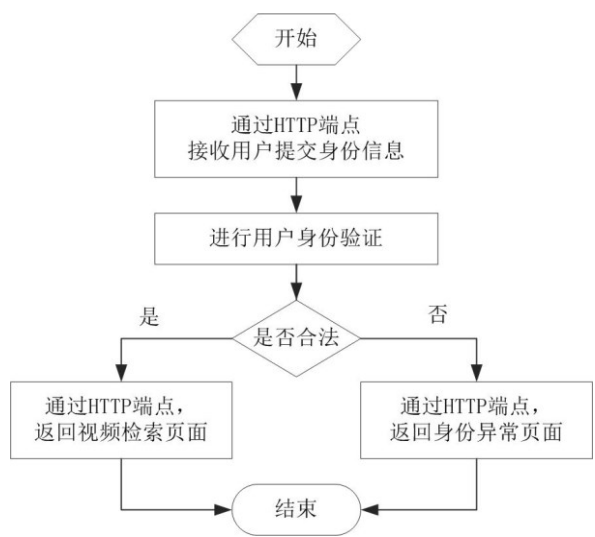

图3 用户身份验证流程图

## 2.2.2 视频资源检索

**Keywords** 通过身份验证后的用户,进入视频资源检索页面,用户输入或者选择查询条件,如淮安市淮阴区,然后提交查询。

**Keywords** 系统收到查询请求,将查询任务按区分服务事务调度算法,调用多个事务处理进程处理。

**Keywords** 事务处理进程解析接收到的查询条件,获取查询参数,动态调用查询语句,从数据库中获取满足条件的视频服务访问调用参数。

**Keywords** 事务处理进程最终将查询结果返回用户。

**Keywords** 用户点击视频列表中摄像机。

**Keywords** 用户端的系统就会将所选择的摄像机详细参数和类型,自动匹配相应的视频解码插件,并连接到相应的摄像机,实现视频自动连接和查看。

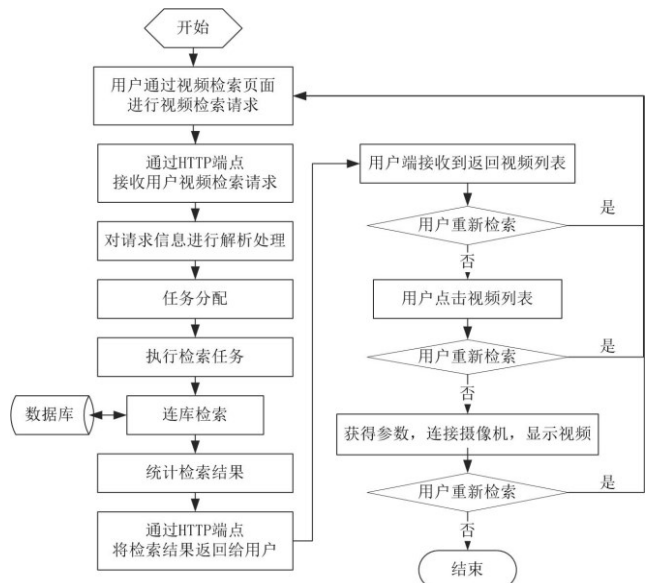

图 4 视频资源检索流程图

## 2.3 关键技术实现

ESB 平台完成了服务发布者与服务请求者之间的连接,在处理大规模数据请求时,为了实现区分服务,解决多个事务请求资源共享问题时,通常采用负载均衡调度机制实现竞争资源的合理分配。

传统的负载均衡任务调度算法有多种,较为常见的有轮询、加权的调度算法。根据人防应急指挥事务处理的特性,参照 WRR 算法[24]的思想,结合 PQ 算法,设计出区分服务事务调度算法,算法的主要思想如下:

本算法为二级调度算法,首先,通过优先级分类器为关键业务提供严格的优先级保证,分成 PQ 队列和 WRR 队列。PQ 队列优先级高于 WRR 队列,用以满足实时性要求比较高的业务流或者比较重要的控制信息流[25];其次,在 PQ 队列和 WRR 队列中设置负载均衡器,定时检测各队列负载率和处理率,根据负载率和处理率的权值,实现事务处理负载均衡[26-28]。

其模型如图 5 所示。

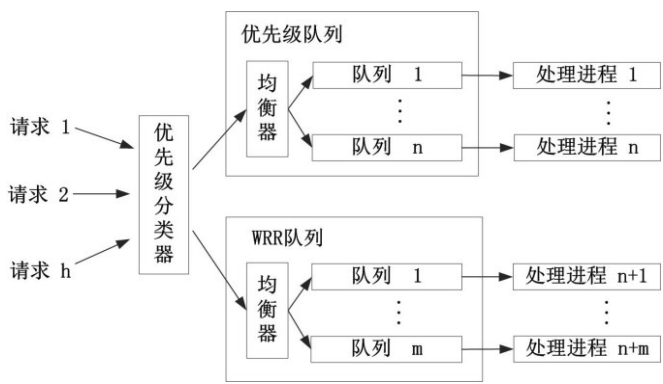

**图 5 区分服务事务调度算法模型图**

### 2.3.1 区分服务事务调度算法

算法中用到的变量

(1) 事务优先级 $P_i$: 表示事务 $i$ 的优先级别,从 0..9,级别值越小,级别越高。

(2) 负载率 $V_i$: 表示当前队列 $i$ 的有待处理的字节数 ($M_i$) 与此队列总空间 ($M_C$) 的比。其中 $M_c$ 为定值,负载率表达式为:

$$V_i = \frac{M_i}{M_c} \quad (1)$$

(3) 处理率 $D_i$: 表示当前队列 $i$ 在 $\triangle T$ 时间内平均处理的字节数。其中 $t_c$ 表示当前计算点的时间,$t_p$ 表示前一个计算点的时间,$\triangle T$ 是 $t_c-t_p$ 时间差值,$M_{tc}$ 是队列 $i$ 在 $t_c$ 时处理的总字节数,$M_{tp}$ 是队列 $i$ 在 $t_p$ 时处理的总字节数,处理率的表达式为:

$$D_i = \frac{M_{tc} - M_{tp}}{\triangle T} \quad (2)$$

在该算法中,优先级分类器根据用户设定的优先级 $P_c$ 对事务的 $P_i$ 进行优先级分类,首先,将优先级大于等于 $P_c$ 的事务放入 PQ 队列,将优先级小于 $P_c$ 的事务放入 WRR 队列,用户可根据实际需要配置 $P_c$ 值;其次,在负载均衡器中进行任务负载分配,PQ 队列和 WRR 队列的负载均衡算法是一样的,具体过程是检测队列中是否有空闲队列,优先分配给空闲队列处理。对于非空闲队列,根据各队列的负载率和前一个处理周期队列的处理率,轮询各队列,将新的任务分配给负载率最小、处理率最大的那个队列。

### 2.3.2 区分服务事务调度算法实现

该算法实现主要过程描述:

(1) 初始化设定系统配置的各项参数,如事务优先级区分值($P_c$)、每个队列总空间大小($M_C$)和队列处理率的统计周期$\triangle T$等。

(2) 当收到一个事务请求时,首先判断该事务优先级 $P_i$,小于等于 $P_c$ 值的分配到 PQ 队列,其它则分配到 WRR 队列处理。

(3) 在均衡器进行任务负载均衡时,先检测 PQ 或 WRR 队列中是否有空闲队列,若有空闲队列,则采用轮询调度算法(RR)优先分配给空闲队列处理。

(4) 没有空闲队列时,动态统计队列的负载率和处理率,根据各非空闲队列的负载率 $V_i$ 和处理率 $D_i$ 进行分配,优先分配给 $V_i$ 小且 $D_i$ 最大值的队列。

(5) 重复步骤(2)、(3)、(4)。

## 3 结束语

整合城市视频资源,提供公共视频服务,是未来智慧城市视频监控系统建设发展的一个方向。ESB 是当今整合异构信息系统,实现数据共享、应用融合的主要技术手段之一。本文提出了泛在视频服务与管理思想,以 ESB 为基础,结合淮安市民防应急指挥调度应用,在信息系统集成领域进行了一些实践,取得了较好的效果,也为今后智慧城市信息化建设做了点有益的尝试。在以后的研究中,我们将在 ESB 的基础上结合语义网技术以及机器学习方法,赋予系统更多的智能[29-32]。

**参 考 文 献**

[1] Shen, F., et al. (2015) SAMAF: Situation Aware Mobile Apps Framework. IEEE International Conference on Pervasive Computing and Communication Workshops, St. Louis, 23-27 March 2015, 26-31

[2] Vaka, P., et al. (2015) PEMAR: A Pervasive Middleware for Activity Recognition with Smart Phones. IEEE Interna

tional Conference on Pervasive Computing and Communication Workshops, St. Louis, 23-27 March 2015, 409-414. https://doi.org/10.1109/PERCOMW.2015.7134073
[3] Shen, F. (2012) Situation Aware Mobile Apps Framework. Master Thesis, University of Missouri, Kansas City. https://mospace.umsystem.edu/xmlui/handle/10355/15637
[4] Shen, F. (2015) A Pervasive Framework for Real-Time Activity Patterns of Mobile Users. Pervasive Computing and Communication Workshops (PerCom Workshops), 2015 IEEE International Conference on, St. Louis, 23-27 March 2015, 248-250.
[5] 曾文英，赵跃龙，齐德昱.ESB原理、构架、实现及应用 计算机工程与应用 2008,44（25） 225-228
[6] Psiuk, M., et al. (2012) "Enterprise service bus monitoring framework for SOA systems." IEEE Transactions on Services Computing 5.3 (2012): 450-466. https://doi.org/10.1109/TSC.2011.32
[7] Dasgupta, S., et al. (2014) SMARTSPACE: Multiagent Based Distributed Platform for Semantic Service Discovery. IEEE Transactions on Systems, Man, and Cybernetics: Systems, 44, 805-821. https://doi.org/10.1109/TSMC.2013.2281582
[8] Chen, Z., et al. (2013) Collaborative Mobile-Cloud Computing for Civil Infrastructure Condition Inspection. Journal of Computing in Civil Engineering, 29, Article ID: 04014066. https://doi.org/10.1061/(ASCE)CP.1943-5487.0000377
[9] Alghamdi, A, et al. (2010) "An interoperability study of esb for c4i systems." Information Technology (ITSim), 2010 International Symposium in. Vol. 2. IEEE, 2010. https://doi.org/10.1109/ITSIM.2010.5561541
[10] Yang, L., et al. (2012) "A Research of Remote Sensing Data Service System Based on Mule ESB." Computer Science & Service System (CSSS), 2012 International Conference on. IEEE, 2012. https://doi.org/10.1109/CSSS.2012.15
[11] Shen, F., et al. (2016) Knowledge Discovery from Biomedical Ontologies in Cross Domains. PLoS ONE, 11, e0160005. https://doi.org/10.1371/journal.pone.0160005
[12] Shen, F. (2016) A Graph Analytics Framework For Knowledge Discovery. PhD Dissertation, University of Missouri, Kansas City. https://mospace.umsystem.edu/xmlui/handle/10355/49408
[13] Shen, F., et al. (2018) BioBroker: Knowledge Discovery Framework for Heterogeneous Biomedical Ontologies and Data. Journal of Intelligent Learning Systems and Applications, 10, 1-20. https://doi.org/10.4236/jilsa.2018.101001
[14] Shen, F., et al. (2018) MedTQ: Dynamic Topic Discovery and Query Generation for Medical Ontologies. arXiv Preprint, arXiv:180203855.
[15] 刘涛，侯秀萍.基于ESB的SOA架构的企业应用研究.计算机技术与发展.Vol.20 No.5 2010.5
[16] Benosman, R., et al. (2013) "Exploiting concurrency for the esb architecture." Engineering of Complex Computer Systems (ICECCS), 2013 18th International Conference on. IEEE, 2013. https://doi.org/10.1109/ICECCS.2013.34
[17] Zhu, Q., et al. (2014) Exploring the Pharmacogenomics Knowledge Base (Pharmgkb) for Repositioning Breast Cancer Drugs by Leveraging Web Ontology Language (OWL) and Cheminformatics Approaches. 19th Pacific Symposium on Biocomputing, Kohala Coast, 3-7 January 2014, 172-182.
[18] Shen, F., et al. (2015) BmQGen: Biomedical Query Generator for Knowledge Discovery. IEEE International Conference on Bioinformatics and Biomedicine, Washington DC, 9-12 November 2015, 1092-1097
[19] Zhang, Y., et al. (2013) An Integrative Computational Approach to Identify Disease-Specific Networks from PubMed Literature Information. IEEE International Conference on Bioinformatics and Biomedicine, Shanghai, 18-21 December 2013, 72-75. https://doi.org/10.1109/BIBM.2013.6732738
[20] 叶忠杰，田文雅，戎成 .ESB在数字校园数据整合中的应用与实践 .计算机时代,2011年第7期 .2011.5.09 .32-34.
[21] 魏铁彬，田凌 .基于企业服务总线的云设计平台的研究与实现 .现代制造工程,2012（8）.1-11.
[22] 彭政 .开源中间件Mule的研究. 科技广场 .2010（5）.17-19.
[23] 薛蕾，蒋朝惠 .基于ESB的智慧城市共享平台设计与实现.计算机技术与发展,2013,第23卷,第3期.218-222.
[24] 高原，顾星，翟明玉，高宗和 .实时监控系统中的任务调度方法研究 .科技资讯 .2012(2) .18-19.
[25] 饶宝乾，侯嘉 .一种基于WRR的新的调度算法 .科学技术与工程 .第11卷，第17期 .2011.6 .4088-4094.
[26] 刘玉艳，沈明玉 .一种LVS负载均衡调度算法WLC的改进 .制造业自动化 .第32卷,第9期 2010.9 .187-191.
[27] 尹德斌，谢剑英 .一种新的加权公平队列调度算法 .计算机工程 .第34卷,第4期 .2008.2 .28-33.
[28] 朱健琛，徐洁，鲁珂 .一种类欧氏距离-负载平衡的云任务调度算法 .计算机仿真 .第29卷，第6期 .2012.6 .159-162.
[29] Tao, C., et al. (2013) Phenotyping on EHR Data using OWL and Semantic Web Technologies. International Conference on Smart Health, Beijing, 3-4 August 2013, 31-32. https://doi.org/10.1007/978-3-642-39844-5_5.
[30] Shen, F., et al. (2014) Using Semantic Web Technologies for Quality Measure Phenotyping Algorithm Representation and Automatic Execution on EHR Data. IEEE-EMBS

International Conference on Biomedical and Health Informatics, Valencia, 1-4 June 2014, 531-534.

[31] Shen, F., et al. (2016) Predicate Oriented Pattern Analysis for Biomedical Knowledge Discovery. Intelligent Information Management, 8, 66. https://doi.org/10.4236/iim.2016.83006

[32] Li, D., et al. (2014) "Towards a multi-level framework for supporting systematic review—A pilot study." Bioinformatics and Biomedicine (BIBM), 2014 IEEE International Conference on. IEEE, 2014. https://doi.org/10.1109/BIBM.2014.6999266